%Paper: astro-ph/9401047
%From: <nowak@chipmunk.cita.utoronto.ca>
%Date: Thu, 27 Jan 94 16:26:49 EST
%Date (revised): Fri, 28 Jan 94 17:05:54 EST

%
% NOTE:  This should TeX much more easily than the previous version
% (otherwise no changes).  Figures available via anonymous FTP
% ftp ftp.cita.utoronto.ca (128.100.76.46), and look in the directory
% /ftp/cita/mike .  The READ_ME file will give the locations of everything.
%
\magnification=1200

 at 10 truept
\font\fourteenrm=cmr10 scaled 1440
\font\twelvebf=cmbx10 scaled 1200

%----------For ninepoint text (quotations)------------------
%-----------(no scriptscripts)------------------------------
\font\ninerm=cmr9
\font\ninei=cmmi9
\skewchar\ninei='177
\font\ninesy=cmsy9
\skewchar\ninesy='60
\font\nineit=cmti9
\font\ninesl=cmsl9
\font\ninebf=cmbx9
\font\ninett=cmtt9
\def\ninepoint{\textfont0=\ninerm \scriptfont0=\sevenrm
              \def\rm{\fam0\ninerm}\relax
              \textfont1=\ninei \scriptfont1=\seveni
              \def\mit{\fam1}\def\oldstyle{\fam1\ninei}\relax
              \textfont2=\ninesy \scriptfont2=\sevensy
              \def\cal{\fam2}\relax
              \textfont3=\tenex \scriptfont3=\tenex
              \def\it{\fam\itfam\nineit}\relax
              \textfont\itfam=\nineit
              \def\sl{\fam\slfam\ninesl}\relax
              \textfont\slfam=\ninesl
              \def\bf{\fam\bffam\ninebf}\relax
              \textfont\bffam=\ninebf \scriptfont\bffam=\sevenbf
              \def\tt{\fam\ttfam\ninett}\relax
              \textfont\ttfam=\ninett
              \setbox\strutbox=\hbox{\vrule
                   height8pt depth3pt width0pt}\baselineskip=11pt
              \adjustlinespacing
              \rm}

%----------For eightpoint text (footnotes)------------------
%-----------(no scriptscripts)------------------------------
\font\eightrm=cmr8
\font\eighti=cmmi8
\skewchar\eighti='177
\font\eightsy=cmsy8
\skewchar\eightsy='60
\font\eightit=cmti8
\font\eightsl=cmsl8
\font\eightbf=cmbx8
\font\eighttt=cmtt8
\def\eightpoint{\textfont0=\eightrm \scriptfont0=\fiverm
                \def\rm{\fam0\eightrm}\relax
                \textfont1=\eighti \scriptfont1=\fivei
                \def\mit{\fam1}\def\oldstyle{\fam1\eighti}\relax
                \textfont2=\eightsy \scriptfont2=\fivesy
                \def\cal{\fam2}\relax
                \textfont3=\tenex \scriptfont3=\tenex
                \def\it{\fam\itfam\eightit}\relax
                \textfont\itfam=\eightit
                \def\sl{\fam\slfam\eightsl}\relax
                \textfont\slfam=\eightsl
                \def\bf{\fam\bffam\eightbf}\relax
                \textfont\bffam=\eightbf \scriptfont\bffam=\fivebf
                \def\tt{\fam\ttfam\eighttt}\relax
                \textfont\ttfam=\eighttt
                \setbox\strutbox=\hbox{\vrule
                     height7pt depth2pt width0pt}\baselineskip=9pt
                \adjustlinespacing
                \rm}

\newdimen\figsizdim \def\figsiz{\dimen\figsizdim}

\expandafter
\ifx\csname draft\endcsname\relax
    \def\psfigcall#1#2{\centerline{\psfig{figure=#1.ps,height=#2,width=#2}}}
\else
    \input landscap
    \def\psfigcall#1#2{\centerline{\psfig{figure=#1.psl,height=#2,width=#2}}}
\fi

\def\insfig[#1,#2,#3]#4.{\midinsert\parindent=0pt
           \figsiz=#2 \divide\figsiz by 1010 \multiply\figsiz by \mag
           \vbox to #3 {\vss\psfigcall{#1}{\figsiz}}
           \def\par{\endgraf\endinsert} \ninepoint
           \ignorespaces {\bf #4.}\quad \ignorespaces}

\def\blackout{\catcode`\\=12\catcode`\{=12\catcode`\}=12\catcode`\$=12
\catcode`\&=12\catcode`\#=12\catcode`\^=12\catcode`\_=12\catcode`\@=12
\catcode`\~=12\catcode`\%=12\catcode`\|=12}

{\obeylines \global\let
                       \relax
    \gdef\fig{\begingroup\obeylines\verbatim}
    \gdef\verbatim{\blackout\def
                   ##1
                   {\setbox0\hbox{##1}%
                    \ifdim\wd0<1pt \endgroup \else\immediate\write1{##1}\fi
                   }
   \string\figcap}}

\def\figcap[#1,#2,#3]#4.{\write1{\null\vfil
           \figsiz=#2 \divide\figsiz by 1010 \multiply\figsiz by \mag
           \vbox to #3{\vss\string\psfigcall{#1}{\figsiz}}}
    \write1{\bigskip\bigskip\centerline{\twelvebf Figure #4}\vfil\string\eject}
    \bigskip\noindent{\bf\ignorespaces#4.}}

\expandafter
\ifx\csname submission\endcsname\relax
      \let\adjustlinespacing\relax
      
      \let\fig=\insfig
\else \def\adjustlinespacing{\baselineskip=1.5\baselineskip}
      \adjustlinespacing
      
      \openout1=captions
\fi

%-----------------General-------------------------------

\parskip \smallskipamount

\def\maybebreak#1{\vskip 0pt plus #1\vsize \penalty -1000
                  \vskip 0pt plus -#1\vsize}

\headline={\ifnum\pageno=0\hfil
           \else\ifodd\pageno\hfil\tenit\rhead\quad\hbox{\tenbf\folio}%
                \else\hbox{\tenbf\folio}\quad\tenit\lhead\hfil\fi
           \fi}
\footline={\hfil}

\def\comment{$\{$\bgroup\eightpoint\aftergroup\endcomment\let\next}
\def\endcomment{$\}$}

%------------Sections and subsections------------------------

\newcount\secnum
\def\section#1\par{\maybebreak{.1}\bigskip\bigskip\bigskip
      \advance\secnum by 1 \subsecnum=0
      \mark{#1} \centerline{\eightrm\the\secnum. \uppercase{#1}}
      \medskip\noindent}

\newcount\subsecnum
\def\subsection#1\par{\ifvmode\maybebreak{.08}\bigskip\fi
      \advance\subsecnum by 1
      \centerline{\the\secnum.\the\subsecnum. \it#1}
      \nobreak\smallskip\noindent}

%------------Numbering equations-----------------------------

\newcount\eqnum
\def\nextenum{\global\advance\eqnum by 1
              \thenumber}
\def\prevenum#1{{\advance\eqnum by -#1 \advance\eqnum by 1
                 \thenumber}\rm}
\def\name#1{\xdef#1{\thenumber}\ignorespaces}
\def\putnum{\eqno(\nextenum)}
\def\thenumber{\rm\the\eqnum}

%-----------------Footnotes---------------------------
\newcount\notenum
\begingroup\catcode`\@=11
\gdef\vfootnote#1{\insert\footins\bgroup\eightpoint
     \interlinepenalty=\interfootnotelinepenalty
     \splittopskip=\ht\strutbox \splitmaxdepth=\dp\strutbox
     \floatingpenalty=20000
     \leftskip=0pt \rightskip=0pt \parskip=1pt \spaceskip=0pt \xspaceskip=0pt
     \smallskip\textindent{#1}\footstrut\futurelet\next\fo@t}

\gdef\note{\global\advance\notenum by 1
    \edef\n@tenum{$^{\the\notenum}$}\let\@sf=\empty
    \ifhmode\edef\@sf{\spacefactor=\the\spacefactor}\/\fi
    \n@tenum\@sf\vfootnote{\n@tenum}}
\endgroup

%--------------Abbreviations---------------------------------

\def\<#1>{{\left\langle#1\right\rangle}}

\def\witchbox#1#2#3{\hbox{$\mathchar"#1#2#3$}}

\def\lsim{\mathrel{\rlap{\lower3pt\witchbox218}\raise2pt\witchbox13C}}
\def\gsim{\mathrel{\rlap{\lower3pt\witchbox218}\raise2pt\witchbox13E}}

\let\twiddle=~ \def~{\ifmmode\tilde\else\twiddle\fi}

\def\frac#1/#2{\mathchoice  {\hbox{$#1\over#2$}} {{#1\over#2}}
               {\scriptstyle{#1\over#2}}{#1/#2}}

\def\pderiv(#1/#2){\mathchoice{\partial#1\over\partial#2}
    {\partial#1/\partial#2} {\partial#1/\partial#2} {\partial#1/\partial#2}}
\def\pderivp(#1/#2){\left(\pderiv(#1/#2)\right)}

\def\tderiv(#1/#2){\mathchoice{d#1\over d#2}{d#1/d#2}
                   {d#1/d#2}{d#1/d#2}}
\def\tderivp(#1/#2){\left(\tderiv(#1/#2)\right)}

%----------------References-----------------------------
\def\apj{ApJ}

\def\apjs{ApJ Suppl}
\def\aj{AJ}
\def\mnras{MNRAS}

\def\annrev{ARA\&A}

\def\nat{Nature}

\def\rhang{\noindent\hangindent\parindent\hangafter1}
% journals, use: \rj{names year}{journal}{volume}{page}
\def\rj#1#2#3#4{\rhang#1, #2, #3, #4\par}
% For books, use: \rb{authors year}{book name}{press}

% For conf.proceedings, use: \rconf{name year}{book name}{editors}{press}{page}
\def\rconf#1#2#3#4#5{\rhang#1, in #2, ed. #3, #4, p.~#5\par}

%--------------------------------------------------------------
%Commonly used definitions.
% \def\lsim{\lower2pt\hbox{$\buildrel{\scriptstyle <}
% \over {\scriptstyle\sim}$}}
% \def\gsim{\lower2pt\hbox{$\buildrel {\scriptstyle >}
% \over{\scriptstyle\sim}$}}
% \def\lapprox{\lower2pt\hbox{$\buildrel\lower2pt\hbox{${\scriptstyle<}$}
% \over {\scriptstyle\approx}$}}
% \def\gapprox{\lower2pt\hbox{$\buildrel\lower2pt\hbox{${\scriptstyle>}$}
% \over {\scriptstyle\approx}$}}

\def\bbuildrel#1_#2^#3{\mathrel{\mathop{\kern 0pt#1}\limits_{#2}^{#3}}}

\def\etal{{\it et~al.\ }}
\def\eg{{\it e.g.,\ }}

\def\ie{{\it i.e.,\ }}

\def\cf{{\it cf.\ }}

\def\rc{r_{\rm c}}

\pageno=0

\def\psfigcall#1#2{\relax}
\def\insfig[#1,#2,#3]#4.{\relax}

\baselineskip=12pt
\null

\vskip .11\vsize

\centerline{\vbox to .26\vsize{
       \baselineskip=12pt \null\vfil \centerline{\vbox{\halign
       {#\hfil\cr \fourteenrm The Statistics of Gamma-Ray Burst
       Lensing \cr
%             \noalign{\medskip}
%             \fourteenrm Next line.... \cr
%             \noalign{\bigskip}
             \noalign{\bigskip\bigskip}
             Scott A. Grossman and Michael A. Nowak \cr
             \noalign{\medskip}
             Canadian Institute for Theoretical Astrophysics \cr
             McLennan Labs, University of Toronto \cr
             60 St.~George Street \cr
             Toronto, Ontario  M5S 1A7 \cr
             Canada \cr
             \noalign{\medskip} submitted to ApJ \cr
             }
           }} \vfil
}}

\vfil

\noindent{\bf Abstract:}
Possibly the only unambiguous verification that gamma-ray bursts
(GRBs) are at cosmological distances would be the observation of
multiple images of a gravitationally lensed burst.  Each images would
arrive at a different time, but exhibit identical light curves.  We
improve upon previous calculations of GRB lensing by using revised
cosmological burst models based upon better knowledge of burst
spectra, and we consider several sets of cosmological parameters.
Only lensing by the known population of galaxies is considered, and we
compute the lensing rate and amplification bias using three different
lens models.  Assuming that 800 bursts per year are above the
detection threshold of the BATSE experiment on the Compton Gamma-Ray
Observatory, we predict one lensing event every 1.5--25 years, with a
median time delay between images of about 7 days.  However, BATSE has
an instrumental efficiency of about 12\% for detecting double bursts,
decreasing the detection rate to fewer than one per century for some
cosmological parameters.  This inefficiency can be overcome partially
if cases of quadruple imaging by elliptical potentials are common, but
simulations show that lensing by elliptical potentials enhances the
detection rate by only about 20\%.  We conclude that it is unlikely
that the BATSE experiment will detect a lensed burst, but that a
future experiment, which is either much more sensitive or has a much
more efficient duty cycle, could.

\vfil \eject \def\lhead{GRB Lensing}
\let\rhead=\lhead
\def\zl{z_{\rm l}}
\def\zs{z_{\rm s}}
\def\zmax{z_{\rm max}}
\def\kev{\rm keV}
\def\kpc{\rm kpc}
\def\kms{\,\rm km\,s^{-1}}
\def\rs{r_{\rm s}}
\def\vrat{\langle V/V_{\rm max}\rangle}

\section Introduction

Data collected by the Burst and Transient Source Experiment (BATSE) on
the Compton Gamma-Ray Observatory (CGRO) have generated interest in
the cosmological origin scenario for gamma-ray bursts (GRBs) (see
Blaes 1994 for a review of various scenarios).  In particular, the
isotropic distribution of bursts is not compatible with any known
galactic distribution (disk or halo) of objects (Mao \& Paczy\'nski
1992b).  The GRBs must originate beyond $18\,\kpc$ from the galactic
center or be part of the solar system (\eg in the Oort cloud) to have
the required isotropy.  Unfortunately, the flux distribution of bursts
alone cannot distinguish among the local, extended halo, and
cosmological possibilities (Lubin \& Wijers 1993).

Since BATSE's spatial resolution ($\sim 4^\circ$) is insufficient to
unambiguously associate individual bursts with individual high
redshift galaxies, perhaps the only definitive proof of the
cosmological origin of GRBs that BATSE could provide is a detection of
a gravitational lensing event.  A lensed GRB would show multiple
images, but unlike quasars, they would not be identified as a
spatially resolved pair (or quadruplet).  Rather, they would appear as
two bursts that arrive at different times and are located within the
same positional error box on the sky.  They will exhibit the same
light curve, but probably will have different amplitudes.

Since the first suggestion that GRBs might be lensed (Paczy\'nski
1986), several authors have presented detailed calculations of GRB
lensing.  Mao (1992b) considered lensing by the known population of
galaxies.  He found that GRBs should be lensed at a rate of about one
per thousand, which is comparable to the rate of quasar lensing
(Ostriker \& Vietri 1986). Other authors have considered lensing by a
significant cosmological density of point masses (Mao 1992b; Blaes and
Webster 1992).  Computations of GRB microlensing (Paczy\'nski 1987;
Mao 1993) and even ``femtolensing'' (Gould 1992) have been considered.
Narayan and Wallington (1992a) focused on what can be learned about
the lens itself if a lensed burst is observed.  Nemiroff \etal (1993)
have examined the first 44 GRBs detected by BATSE for evidence of
lensing. The absence of any lensed bursts in this sample is used to
set rather loose constraints upon the density of $10^6$--$10^8M_\odot$
objects in the universe.

Considering the predicted rate of GRB lensing, it is not surprising
that no lensing events were found in the small sample examined by
Nemiroff \etal (1993).  Indeed, if only galaxies act as lenses, none
should be present in the sample of more than 300 bursts catalogued so
far.  Furthermore, even if there were a lensing event in the
catalogued data, Wambsganss (1993) has claimed that the amount of
noise present in the BATSE light curves may overwhelm any hope of
identifying it.  Nowak and Grossman (1994) find that this is
excessively pessimistic.  They conclude that a large fraction of
bursts that are bright enough to trigger the BATSE detector
($5.5\sigma$ above the background noise) are also bright enough to
have light curves that can be distinguished from other distinct
bursts.  Since previous calculations suggest that in a few more years
of data there should be one lensing event if GRBs are cosmological,
searching for lensed bursts may be an exercise with a potentially
large return.

The purpose of this paper is to revise and expand upon the
computations of lensing by the known population of galaxies.  In
particular, we find a significantly lower lensing rate than did Mao
(1992b), for two important reasons.  First, he used a cosmological
burst model where bursts can be seen to redshifts farther than are
likely.  According to recent models using improved burst spectra
(Fenimore \etal 1993; Tamblyn \& Melia 1993, hereafter TM93), bursts
are not likely to be seen beyond $z\sim 1$, whereas Mao (1992b) used a
model where they can be seen to $z\sim 1.5$ (Mao and Paczy\'nski
1992a).  Second, Mao did not use the appropriate selection criterion
for the occurrence of a lensing event, namely whether or not the
second brightest burst is observable.  In \S2 we construct
cosmological models of GRBs using the improved spectrum of TM93 for
cosmologies $(\Omega,\Lambda)=(1.0,0)$, $(0.1,0)$, $(0.1,0.9)$.  In
\S3 we use these three cosmological models to compute the rate of
observable double bursts, using the population of known galaxies,
modeled as singular isothermal spheres, as lenses.  In addition we
consider the lensing rate if the lenses are non-singular isothermal
spheres or have finite mass and no dark, isothermal halos.  The
lensing rate is high enough that a lensing event has a significant
chance of occurring during BATSE's lifetime.  We discuss how BATSE's
instrumental inefficiency for detecting lensing events defeats this
moderately optimistic rate.  In \S4 we consider lensing by an
elliptical potential, and derive from simulations the probability for
multiple imaging with two, three, or four observable components.  If
four component bursts are sufficiently common, then the instrumental
inefficiency can be partly overcome.  We demonstrate that lens
ellipticity enhances the detection rate of lensing events by a small
amount.  We summarize and discuss the implications of our results for
the BATSE experiment and potential future experiments in \S5.

\section The GRB Population

Analysis of the flux distribution of GRBs (Meegan \etal 1992) shows
that the observed $\vrat$ statistic (Schmidt 1968) is not compatible
with a homogeneous distribution.  (This statement presupposes that the
flux decreases as the inverse square of the distance.)  Although the
$\vrat$ statistic can be reconciled with a local population that is
absent at great distances, the observed GRB isotropy is inconsistent
with a local galactic halo population (unless the core radius of the
distribution is anomalously large, Mao \& Paczy\'nski 1992b) and with
a local population associated with the Oort cloud, Maoz 1993, Clarke
\etal 1994).  Another possibility is that the inverse square law does
not hold, as would be the case if bursts are at cosmological distances
($z\gsim 0.3$) such that the curvature of the universe is important.
Several authors (Dermer 1992; Mao \& Paczy\'nski 1992a; Piran 1992;
Wickramasinghe \etal 1993; Fenimore \etal 1993; TM93) have shown that
such departures from Euclidean geometry can account for the GRB flux
distribution.  The observation that fainter bursts have longer
duration (Norris \etal 1994) is consistent with cosmological time
dilation of distant bursts, and supports the cosmological scenario.

We assume GRBs originate at cosmological distances, and that the
population can be described by standard candle luminosity $L_0$ and a
constant burst rate per comoving volume $n_0$.  The maximum distance
to which a burst can be seen measures the departure from Euclidean
geometry, and thus, $L_0$ is chosen to reproduce the observed $\vrat$.
Based on 207 bursts in the BATSE burst catalog (Fishman \etal 1992) up
to March 5, 1992 (overwrites are excluded), we compute $\vrat=0.32$,
in agreement with that from the slightly smaller sample in Meegan
\etal (1992).  The burst rate per unit volume $n_0$ is chosen to
reproduce the observed burst rate, $N\approx 800\,\rm yr^{-1}$.

The details of this calculation depend on the GRB spectrum.  We adopt
the double power-law model of TM93, which is based on the analysis of
burst spectra by Band \etal (1993).  The photon spectrum (\ie photons
per unit time per unit energy) is described by $$S_E(E)=\cases{
AE^{\alpha_1-1}, & $E\le E_b$\cr
AE_b^{\alpha_1-\alpha_2}E^{\alpha_2-1}, & $E\ge E_b$,\cr
}\putnum$$\name\spectrum where $E_b=300\,\kev$ is the break in the
power-law slope.  The power law exponents are $\alpha_1=1$,
$\alpha_2=-1$.  The coefficient $A$ is calibrated by equating the
energy emitted in the bandpass $50$--$300\,\kev$ (the bandpass in
which BATSE bursts are observed) to the restframe luminosity, $L_0$.

The photon flux observed in the $50$--$300\,\kev$ bandpass depends on
the distance to the burst, measured by its redshift $\zs$, and is
given by $$\phi(\zs)={S(\zs)\over{4\pi d_L(\zs)^2}},\putnum$$
\name\photflux where $d_L(\zs)$ is the luminosity distance and
$S(\zs)$ is the K-corrected photon luminosity given by
$$S(\zs)=\int_{50\kev}^{300\kev}S_E[(1+\zs)E]\,dE,\putnum$$\name\kphot
that is, the photon emission rate corrected for the redshifted
bandpass.  Evaluating this integral, we find $$S(\zs)=\cases{
\displaystyle L_0\,4.3\times 10^6(1+\zs)^{-1}\left[2-(1+\zs)^{-1}
-{1\over 6}(1+\zs)\right], &$\zs\le 5$\cr
\displaystyle L_0\,4.3\times 10^6(1+\zs)^{-2}5, &$\zs\ge 5$\cr
}\putnum$$\name\kphoti where $L_0$ is measured in $\rm erg\,s^{-1}$.
Bursts with fluxes that are $5.5\sigma$ above background in either the
$64\,\rm ms$, $256\,\rm ms$, or $1024\,\rm ms$ time bins trigger
BATSE.  The $1024\,\rm ms$ trigger threshold is most sensitive, so we
adopt its flux limit, $\phi_{\rm min}=0.3\,\rm
photon\,s^{-1}\,cm^{-2}$ for $\gsim50\%$ detection probability
(Fishman \etal 1992).  The flux limit $\phi_{\rm min}$ defines the
maximum redshift $\zmax$ out to which bursts can be seen.

The rate at which we observe GRBs that originate in redshift interval
$[\zs,\zs+d\zs]$ is given by
$${dN\over{d\zs}}={n_0\over(1+\zs)}{dV\over{d\zs}}.\putnum$$\name\dndz
The factor $1+\zs$ comes from the gravitational time dilation of the
burst rate, and $dV/d\zs$ is the comoving volume element.  Using
equations (\photflux) and (\dndz), we can calculate how many bursts
are observed with flux $\phi$.  The theoretically predicted $\vrat$
for this flux distribution is given by $$\left\langle{V\over{V_{\rm
max}}}\right\rangle=
\left\langle\left({\phi\over{\phi_{\rm min}}}\right)^{-3/2}
\right\rangle={\int_0^{\zmax}
\left(\phi(\zs)/\phi_{\rm min}\right)^{-3/2}(dN/dz_s)\,d\zs
\over{\int_0^{\zmax}(dN/d\zs)\,d\zs }}.\putnum$$\name\thvrat

The only missing ingredient required to carry out this computation is
the specific choice of a cosmological model for the comoving volume
element and the luminosity distance $d_L$.  We will consider three
Friedmann-Robertson-Walker cosmologies described by density and
cosmological constant parameters $(\Omega,\Lambda)$ given by
$(1.0,0)$, $(0.1,0)$, $(0.1,0.9)$.  Cosmologies with cosmological
constant much more than 0.9 probably already are excluded by the
quasar lensing rate (Fukugita \etal 1992).  The proper, luminosity,
and angular distances are given by (\cf Carrol, Press, \& Turner 1992)
$$d_M(\zs)={c\over{H_0}}\int_0^{\zs}\left[(1+z)^2(1+\Omega z)
-z(2+z)\Lambda\right]^{-1/2}\,dz,\eqno(\nextenum a)$$\name\dm
$$d_L=(1+\zs)d_M,\eqno(\prevenum1 b)$$
$$d_A=d_M/(1+\zs).\eqno(\prevenum1 c)$$ In general, these distances
must be computed numerically. The angular distance between two
redshifts $d_A(\zl,\zs)$ can be found by replacing the lower bound of
integration in equation (\dm a) with the lower redshift.  The comoving
volume element is given by $${dV\over{d\zs}}=4\pi
d_M^2{d\over{d\zs}}d_M=4\pi d_M^2{c\over{H_0}}
\left[(1+\zs)^2(1+\Omega \zs)-\zs(2+\zs)\Lambda\right]^{-1/2}.\putnum$$
\name\dvdz
The quantitative effects of using an alternative, clumpy universe
cosmology (Dyer \& Roeder 1973) are negligible for the moderate
redshifts $\zs\sim 1$ out to which bursts can be seen (Fukugita \etal
1992).

The values of $L_0$, $n_0$, and $\zmax$ that reproduce the observed
$\vrat$ and burst rate are listed in Table 1.  The parameters for
$\Omega=1$, $\Lambda=0$ are in approximate agreement with those of
Fenimore \etal (1993) (presumably the differences are a consequence of
using a different spectrum).  Our $\Omega=0.1$, $\Lambda=0.9$ model
also is consistent with that of Piran (1992), but no prior results are
available for the $\Omega=0.1$, $\Lambda=0$ cosmology.  We expect that
a more careful analysis, for example including PVO data, would modify
the results only slightly.  We have not been more precise since burst
modeling is not the aim of this paper, but is only one ingredient in
the lensing calculation to follow.

In Figure 1, the redshift distributions of bursts are shown.  In fact,
for each cosmology two curves are drawn.  One curve is the computation
of equation (\dndz), and the second is the same computation, but
showing the incompleteness as fluxes approach and fall below
$\phi_{\rm min}$.  We have found that the data in the BATSE efficiency
table for $1024\,\rm ms$ triggers (Fishman \etal 1992) is fit very well by
the function $$\epsilon(\phi)=\cases{
\displaystyle 0, &$\phi\le \phi_0$\cr
\displaystyle 1-e^{-(\phi-\phi_0)/\alpha}, &$\phi\ge \phi_0$\cr
}\putnum$$\name\eps with $\phi_0=0.21\rm\,photon\,s^{-1}\,cm^{-2}$ and
$\alpha=0.105\rm\,photon\,s^{-1}\,cm^{-2}$.  We use equation (\eps) as
the detection efficiency in this and all subsequent calculations.  The
dotted vertical lines are drawn at the values of $z_{\rm max}$, where
by definition the incompleteness is 50\%.  Note that some bursts
originating at $\zs>\zmax$ will be seen when Poisson uncertainty in
the photon rate brings the measured flux above $\phi_{\rm min}$.

\fig[PS/dndz,\hsize,.6\hsize] 1.
The redshift distribution of GRB as computed from equation (\dndz),
with and without correction for incompleteness at faint fluxes near
the detection threshold.  The curves are labeled by the cosmological
parameters $(\Omega,\Lambda)$.  The dotted vertical lines indicate the
values of $\zmax$, which by definition, correspond to 50\%
completeness.  The integrals under the curves allowing for
incompleteness give 800 bursts per year.

Band (1992, see also Petrosian 1993) has argued that failure to
account for incompleteness near the flux threshold invalidates many
comparisons between the observed and predicted $\vrat$.  We test the
sensitivity of our parameters to incompleteness by including the
detection efficiency function, equation (\eps) in our calculation of
$\langle{V\over{V_{\rm max}}}\rangle$.  Multiplying the top and bottom
integrands of equation \thvrat by the detection efficiency and
integrating the bursts between $z=0$ and $z=\zmax$, we find
$\vrat=0.27$ instead of 0.32 for our $\Omega=1$, $\Lambda=0$
parameters.  Furthermore, our and other models (\eg TM93) reproduce
the observed $\vrat$ at thresholds well above the regime where
incompleteness is relevant, but where cosmological effects are still
important.  Whether or not we include this incompleteness should
modify the details of our cosmological model only slightly.

\section The Rate of Lensing by Galaxies

In this section we consider lensing by circular lens models.  (See
Blandford \& Narayan 1992, Narayan \& Wallington 1992b for reviews of
the physics of lensing.)  Such lenses alway produce at least one image
of magnification $M_1$ greater than unity.  This image is usually the
brightest and most distant from the center of the lens.  If there are
multiple images, a second image with magnification $M_2$ appears on
the opposite side of the lens.  This image can be very bright or very
faint, depending on the lens model and the position of the source with
respect to the lens.  If the lens has a non-singular core, there can
be a third image of magnification $M_3$ in the core.  This image is
almost always very faint.

The differential lensing rate of GRBs is given by $${dN_{\rm
l}\over{d\zs}}={dN\over{dz_s}}{\sigma_{\rm
tot}(\zs)\over{4\pi}},\putnum$$\name\dnldz where $\sigma_{\rm
tot}(\zs)$ is the total cross section of all lenses in the sky to
produce an observable multiple burst from a source in the interval
$[\zs,\zs+d\zs]$.  If galaxies are the lensing masses, we can write
the total cross section as $$\sigma_{\rm tot}(\zs)=
\int_0^{\zs}d\zl{dV\over{d\zl}}\int_0^\infty dL\,\Phi(L)\sigma(\zl,\zs,L).\
\putnum$$\name\sigtot
The cross section for a single galaxy, $\sigma(\zl,\zs,L)$, depends on
the lens mass, which is directly related to lens luminosity.
Weighting these cross sections by the galaxy luminosity function
$\Phi(L)$, the inner integral sums over all lens masses at a single
lens redshift $\zl$, and the outer integral sums over all the lensing
galaxies at each lens redshift.  The lens population is characterized
by the Schecter luminosity function,
$$\Phi(L)dL=\Phi^*e^{-L/L_*}\left({L\over{L_*}}\right)^\beta{dL\over{L_*}},
\putnum$$\name\lfun
with parameters $\Phi^*=1.56\times 10^{-2}h_{100}^3\,\rm Mpc^{-3}$,
$\beta=-1.1$, as assumed in Fukugita and Turner (1991, hereafter
FT91).  The dimensionless Hubble constant is defined by
$h_{100}=H_0/100\,\rm km\,s^{-1}Mpc^{-1}$.

The lensing cross section of a single galaxy of luminosity $L$ to
produce an {\it observable} multiple burst is
$$\sigma(\zl,\zs,L)=\int_0^{r_{\rm crit}}
d\rs\,2\pi\rs\epsilon[M_2(\rs)\phi(\zs)].\putnum$$\name\cross The
variable $\rs$ is the angular distance of a source to the lens axis.
We have assumed we can relate the luminosity of a lens to its lens
parameters.  The value of $r_{\rm crit}$, the outermost source radius
that gives multiple images, depends on the particular lens model.
Whether or not multiple images will be seen depends on the
magnification $M_2$ of the fainter image with photon flux
$M_2\phi(\zs)$.  If the detection efficiency $\epsilon[M_2\phi(\zs)]$
is below unity, the effective cross section is reduced below its
maximum value $\pi r_{\rm crit}^2$.  (We note that Mao, 1992b assumed
that the full isothermal cross section contributed.  Blaes \& Webster,
1992 did use the right selection criterion in their computation for
point lenses.)  Depending on the source position $\rs$, $M_2$ can be
greater or less than unity.  If $M_2$ is greater than unity, it is
possible to see bursts beyond $\zmax$, but if it is less than unity,
bursts may have to be considerably closer than $\zmax$ for two images
to be detectable.

In addition to the rate at which GRBs are lensed, we would like to
know the characteristic time interval between the arrival of two
images.  The distribution of the time delay $\Delta t$ between two
bursts, which was computed first by Mao (1992b), is given by the
formula $${dN_l\over{d\Delta t}}=\int_0^\infty d\zs{dN\over{d\zs}}\,
\int_0^{\zs} d\zl{dV\over{d\zl}}\int_0^\infty dL\,\Phi(L)
{\partial\sigma(\zl,\zs,L)\over{\partial\Delta t}}
\epsilon[M_2(\rs)\phi(\zs)].\putnum$$\name\tdist
For any particular lens model, the time delay $\Delta t$ can be
related to the source position $\rs$, so that the differential cross
section is $${\partial\sigma\over{\partial\Delta t}}={2\pi\rs
d\rs\over {\partial\Delta t\over{\partial\rs}}d\rs}.\putnum$$
\name\diffcross

\subsection The Singular Isothermal Sphere (SIS)

The singular isothermal lens (SIS) produces two images of the source
whenever the source position lies inside the Einstein radius,
$\rs<r_E$, given by $$r_{\rm
crit}=r_E=4\pi{v^2\over{c^2}}{d_A(\zl,\zs)\over{d_A(0,z_s)}},\putnum$$
\name\rcrit
where $v$ is the line-of-sight velocity dispersion of the lensing
galaxy.  It is related to the lens luminosity via the Faber-Jackson or
Tully-Fisher relations,
$${L\over{L_*}}=\left({v\over{v_*}}\right)^n.\putnum$$ Following FT91,
the lens population is assumed to be comprised of E, S0, and S
galaxies in the proportion 12\%, 19\%, and 69\%, with corresponding
parameters $v_*=225\sqrt{3/2}\kms$, $n=4$ for Es,
$v_*=206\sqrt{3/2}\kms$, $n=4$ for S0s, and $v_*=144\kms$, $n=2.6$ for
Ss.  The factor of $\sqrt{3/2}$ in the velocity dispersion of Es and
S0s accounts for isothermal dark halos surrounding more compact mass
distributions (Turner, Ostriker, \& Gott 1984; FT91).

In the SIS lens, the magnification $M_2$ is related to the source
position by $$M_2=r_E/\rs-1.\putnum$$\name\magtwo For $\rs>r_E/2$
(three quarters of the maximum cross section $\pi r_E^2$), $M_2$ is
less than unity, and $M_2$ goes to zero as $\rs$ approaches $r_E$,
making the second image unobservable.  In no case does a galaxy
contribute the maximum cross section $\pi r_E^2$.

Combining equations (\dnldz)--(\cross), (\rcrit)-(\magtwo) and
evaluating the integrals numerically, we compute the lens rate as a
function of source redshift $\zs$ for each cosmology.  The results are
shown in Figure 2.  For ``standard'' cosmological values $\Omega=1$,
$\Lambda=0$, we get a total multiple burst rate of $0.09\,\rm
yr^{-1}$.  This is quite a bit smaller than the value of $1.1\,\rm
yr^{-1}$ predicted by Mao (1992b), both because he adopted a GRB model
in which unlensed bursts could be seen out to redshifts $\zmax\sim
1.5$ and because he used the maximum isothermal cross section.  As
expected (\cf Carrol, Press, \& Turner 1992; Turner 1990; Fukugita,
Futamase, \& Kasai 1990), the finite $\Lambda$ model gives the largest
lensing rate, about 7 times larger than the standard model.

\fig[PS/sisrate,\hsize,.6\hsize] 2.
The differential lensing rate as a function of source redshift $\zs$
for the SIS lens model.  The curves are labeled by the cosmological
parameters $(\Omega,\Lambda)$, and the cumulative lensing rates $N_l$
(\ie the integrals under the three curves) are given at upper right.

The amplification bias is the ratio of the number of lensed bursts to
the lensing rate computed as if bursts were not magnified.  That is,
it is the increase in the number of lensing events because sources
intrinsically fainter than the flux threshold $\phi_{\rm min}$ are
counted.  We compute the hypothetical, no magnification lensing rate
by replacing the detection frequency in equation (\cross) with
$\epsilon[\phi(\zs)]$, \ie assuming the magnification is unity.  The
cumulative burst rates and the amplification biases computed for the
various cosmologies are presented in Table 2.  The amplification
biases are smaller than for quasar lensing.  Two factors contribute to
this.  One is that the observed GRB flux distribution is not as steep,
so that the increase in the number of observable sources resulting
from amplification bias is not as large.  The other, equally important
factor is that amplification bias is computed using the magnification
probability of the fainter second image.  This differs from most
computations of quasar lensing, where the magnification probability is
computed either for the brighter image or for the total magnification
(\eg Turner, Ostriker \& Gott 1984; FT91; but see Sasaki \& Takahara
1993 for an exception).  In either case, the magnifications of the
quasars are always greater than unity, so that there is no loss of
cross section for sources at $z<\zmax$.  For GRBs, however, we find a
bias $B\sim 1$ for the $\Omega=1$, $\Lambda=0$ model.  Even though
lensed GRBs can be seen to a somewhat higher redshift than unlensed
bursts (compare Figs. 1 and 2), the loss of lensing cross section for
small $M_2$ almost perfectly offsets this increase.  That is, even for
sources at low redshifts, the second image will be too faint to see
for some source positions that yield double images.  In fact, biases
can be less than unity (Sasaki \& Takahara 1993), as in the
$\Omega=0.1$, $\Lambda=0$ model.

Even if our models of GRBs are more or less correct, reproducing the
right $\zmax$, there are uncertainties in the lens rate resulting from
uncertainties in the lens parameters.  In particular, since the lens
rate varies as $N_l\sim\Phi^*v^4\sim\Phi^*v_*^4(L/L_*)^{4/n}$, the
10\% uncertainty in $L_*$, 22\% uncertainty in $\Phi^*$, and 10\%
uncertainty in $v_*$ (\eg FT91; Mao 1992a) give nearly a 50\%
uncertainty in the computed lensing rates.  Using a different
observational luminosity function, Wallington and Narayan (1993) find
a lensing rate about 40\% larger than that given by the FT91
parameters.

We next compute the statistical distribution of time delays.  The time
delay of a pair of images in a lens of Einstein radius $r_E$ is given
by $$\Delta t=f(\zl,\zs,r_E)\rs,\hskip20pt \rs<r_E\putnum$$\name\deltt
where we define the function $$f(\zl,\zs,r_E)=2(1+\zl){d_A(0,\zl)\over
c} {d_A(0,\zs)\over{d_A(\zl,\zs)}}r_E.\putnum$$ Then the differential
cross section (eq. \diffcross) is
$${\partial\sigma\over{\partial\Delta t}}={2\pi\Delta
t\over{f^2}}.\putnum$$\name\diffcrosst The corresponding source
position is given by $\rs(\Delta t)=\Delta t/f$, and hence the
magnification $M_2$ of the second image can be calculated by equation
(\magtwo).

Substituting equation (\diffcrosst) into equation (\tdist), we compute
the distribution of time delays.  The results are shown in Figure 3.
The curves are divergent at small time delay due to the infinite
number of low luminosity sources predicted by the luminosity function
(eq. \lfun), but the total lensing rate, \ie the integral under the
curves, is finite and gives, of course, precisely the same cumulative
lensing rate as the calculation for Figure 2.  The median time delay
is about $0.02\,h_{100}^{-1}\,\rm yr$ for the two $\Lambda=0$
cosmologies and about $0.04\,h_{100}^{-1}\,\rm yr$ for the
$\Omega=0.1$, $\Lambda=0.9$ cosmology.

\fig[PS/tdist,\hsize,.6\hsize] 3.
The lensing rate as a function of time delay $\Delta t$ for the SIS
lens model.  The curves are labeled by the cosmological parameters
$(\Omega,\Lambda)$, and the cumulative lensing rates $N_l$ (\ie the
integrals under the three curves) are given at upper right.  The
median time delays are about $0.02\,h_{100}^{-1}$ years (7 days).

\subsection The Non-Singular Isothermal Sphere (NSIS)

Realistic galaxy lenses probably have finite core radii.  As is
well-known, circular lenses with finite core radii can, in addition to
the pair of images of a singular lens, produce a third image near the
core of the lens.  We adopt the usual non-singular isothermal lens
model (NSIS, \eg Hinshaw \& Krauss 1987), for which we can calculate
image positions and magnifications for any source position.  The only
difference between this calculation and the previous one is that we
have replaced equation (\magtwo) with the appropriate relation for an
NSIS lens.

In previous investigations of quasar lensing (FT91; Mao 1992a), the
importance of the core radius in galaxy lenses was based upon the
observations of 42 elliptical galaxies by Lauer (1985).  Lauer
observed that 14 ellipticals had resolved cores, with half light radii
that approximately follow the relation
$${r_{1/2}\over{r_{1/2}^*}}\approx\left({L\over{L_*}}\right)^{1.2},\putnum$$
\name\rcscale
where $r_{1/2}^*d_A(\zl)=0.16\,\kpc$.  The core radius, $\rc$, of the
lens model is related to the half light radius by
$r_{1/2}=\sqrt{3}r_c$.  The remaining ellipticals had unresolved
cores, so that we may regard the above values of $r_{1/2}$ as upper
limits.  We assume all types of galaxies (E, S0, S) have core radii
following the scaling of equation (\rcscale).  The resulting lensing
rates are reduced by about a factor of two, and this may be an
overestimate if galaxy cores are often smaller than the above
prediction.  The lensing rates and amplification biases for the NSIS
model are listed in Table 2.

To compute how often the third core image is seen, we simply replace
the magnification $M_2$ in the detection efficiency computation with
the magnification of the core image $M_3$.  We find that the third
image will be observable in fewer than 1\% of the double bursts.  The
third image will be visible only in the very rare, very luminous
lenses where the core radius is large and the central image is only
moderately demagnified.

\subsection A Finite Mass Lens Model (QDEV)

If galaxies, particularly ellipticals, do not have massive isothermal
halos, lensing computations based on a density profile containing
finite mass may be more appropriate.  For this analysis we adopt the
quasi-de Vaucouleurs (QDEV) profile of Grossman and Saha (1994).  This
lens model is characterized by the deflection law
$$\alpha(r)={4GM\over{d_A(0,\zl)c^2}}{r\over{(\rc+r)^2}}.\putnum$$
This deflection law closely resembles the true de Vaucouleurs
deflection law (Sanitt 1976) if the core radius $\rc=0.4r_e$, where
$r_e$ is the half mass radius, but it is simpler to use in numerical
calculations.

We use the relations of Maoz and Rix (1993) to relate the luminosity
of a lens to the lens parameters, core radius and mass.  These
relations are
$$\rc=0.4r_e=0.4r_e^*\left({L\over{L_*}}\right)^a\putnum$$ and
$${M\over L}=\left({M\over L}\right)^*\left({L\over{L_*}}\right)^b,
\putnum$$ where the parameters
are $r_e^*=4\kpc$, $(M/L)^*=10$, $a=1.2$, $b=0.25$.  This lens can
give three images, and as with the NSIS model, we are able to compute
their positions and magnifications for any source position.  Using the
magnification of the second image $M_2$ in the calculation of
detection efficiency, we find the lensing rate of this model.

The lensing rate of GRBs, assuming all types of galaxy lenses can be
modeled in this way, is given in Table 2.  These rates are generally
about half the SIS rate.  The third core image is bright enough to be
seen only about $0.1\%$ of the time.  The amplification biases are
quite small.  This is essentially because the cross section for
multiple imaging is much larger than for the SIS lens (approaching the
formally infinite cross section for a point lens), but the cross
section for two images to be bright enough to observe is similar to
the SIS cross section.  Although lensing permits sources somewhat
fainter and more distant than $\zmax$ to be observed, the loss of
lensing cross section associated with small values of $M_2$ more than
offsets this increase.

The lensing rate of this model scales as
$N_l\sim\Phi^*(M/L)L\sim\Phi^*(M/L)^*/L_*^b$.  The 22\%, 20\%, and
10\% uncertainties in $\Phi^*$, $(M/L)^*$, and $L_*$ (FT91; Maoz \&
Rix 1993) combine to give about a 30\% uncertainty in the computed
lens rate.

\subsection The BATSE Detection Rate

The most pessimistic rate of observable double bursts in Table 2 is
one every 25 years, while the most optimistic is one every 1.5 years.
A standard $\Omega=1$, $\Lambda=0$ cosmology gives one every 10 years.
Since GRO is likely to operate for at least several years, the odds
are significant that there will be a lensed burst during this period
if GRBs are cosmological.

Unfortunately, BATSE is inefficient at detecting bursts.  Due to Earth
blockage resulting from CGRO's low orbit ($450\,\rm km$, Gehrels \etal
1993), BATSE only ``sees'' about $2.6\pi$ steradians of the sky at one
time, and the detectors have a duty cycle of only about $55\%$
(Fenimore 1993), giving a probability of 34\% (Fishman \etal 1992) to
see any single burst.  Since the time delay between lensed bursts is
typically much longer than the orbital period of the satellite (many
days compared to $\sim$100 minutes), the position of the satellite in
orbit is essentially random with respect to burst arrival times.
Thus, the odds of seeing both components of a double burst are
$34\%\times 34\%=12\%$.  In view of this instrumental inefficiency, if
lenses are of the simple sort described above, it is very unlikely
that multiple bursts will be seen.  This instrumental inefficiency
generally has been ignored in previous GRB lensing papers (an
exception being Mao, 1992b), resulting in undue optimism (\eg Blaes
1994) about the possibility of observing a lensing event in the
lifetime of BATSE.

\section Lensing by an Elliptical Potential

About half of known multiply-imaged quasars show four, rather than
two, bright images (\eg Blandford \& Narayan 1992).  The simplest lens
that can have such an image configuration is one with an elliptical
gravitational potential.  (See Narayan \& Grossman, 1989 for a review
of caustics, critical lines, and images in an elliptical lens.)  If
the requirement to observe a lensing event is for BATSE to detect two
or more images of a quadruple burst, the instrumental detection
efficiency may be much higher than the 12\% of a double burst.  In
general, the probability of detecting two or more images out of $n$
that are bright enough to be seen is given by the binomial probability
$$P(\ge 2|n)=\sum_{i=2}^n \pmatrix{n\cr i\cr} p^i(1-p)^{n-i},\putnum$$
where $p=34\%$.  If only two images are bright enough to be seen, this
gives $P(\ge 2|2)=12\%$; if three, $P(\ge 2|3)=27\%$; and if four,
$P(\ge 2|4)=42\%$.  Thus, if quadruple images are sufficiently common,
then the chance of detecting a lensing event may be better than
predicted for circular lens models.

To investigate the importance of elliptical lenses in GRB lensing, we
model galaxy lenses by elliptical generalizations of the SIS
potential, as used in the statistical lensing of quasars (Wallington
\& Narayan 1993).  The scaled two-dimensional potential $\psi$
(defined such that the deflection angle is $\alpha=-\nabla\psi$) is
given by
$$\psi(x,y)=r_E\left[(1+\epsilon)x^2+(1-\epsilon)y^2\right]^{1/2},\putnum$$
where $\epsilon$ measures the lens ellipticity.  We derive the
statistics of elliptical lenses using a Monte Carlo simulation, where
we randomly distribute isothermal lenses between the observer and
source at $\zs$ according to their luminosity distribution given by
equation (\lfun) and spatial distribution given by the comoving volume
element.  At each source redshift $\zs$ at 0.2 intervals, we randomly
place 10,000 sources within a circle of radius 1.3 times the Einstein
radius (since we are interested only in potentially observable
multiple bursts).  The resulting statistical contribution to the time
delay distribution of a pair of lensed bursts has a weight
proportional to the area of the Einstein ring.

All the images of a single source are located using an efficient grid
search method, where we identify rough image locations on a coarse
grid and iteratively subdivide to higher resolution only those grid
cells which may contain an image.  Each iteration increases the
resolution by a factor of 10, and it goes 8 levels deep.  Sometimes
several neighboring pixels are marked as containing an image, so as a
final step, we compute the $x$ and $y$ components of the lens equation
(\eg $x_s=x-\alpha(x,y)$) at the four corners of each pixel and keep
only pixels where the source position is bounded by the four corner
values.  We use the four corner values of $x_s$ and $y_s$ to
interpolate $x$ and $y$ to a refined image position.

The results for a lens of ellipticity $\epsilon=0.2$ are shown in
Figure 4, where we plot as a bold line the ratio of the cross section
of the elliptical lens for observable multiple imaging,
$\sigma_\epsilon(2+)$, to the cross section of the circular lens for
double imaging, $\sigma_0(2)$.  The cross section
$\sigma_\epsilon(2+)$ is divided into the contributions of two-image
bursts, three-image bursts, and four-image bursts, \ie
$\sigma_\epsilon(2+)=\sigma_\epsilon(2)+\sigma_\epsilon(3)+
\sigma_\epsilon(4)$.  Analyzing the data of Huchra \etal (1983),
Wallington and Narayan (1993) show that few galaxies have
ellipticities exceeding $\epsilon=0.3$.  They find a mean ellipticity
somewhat less than $\epsilon=0.2$, and thus, using $\epsilon=0.2$ in
Figure 4 is roughly representative of a typical galaxy lens.

\fig[PS/elens,\hsize,.6\hsize] 4.
The rate of multiple imaging by an elliptical potential
($\epsilon=0.2$), measured by cross section $\sigma_\epsilon(2+)$,
compared to the rate of double imaging by a circular potential,
measured by cross section $\sigma_0(2)$, is drawn as the solid, bold
line.  The multiple imaging cross section is divided into the
contributions of double, triple, and quadruple images.  (That is, the
bottom curve is $\sigma_\epsilon(2)$, the next one is
$\sigma_\epsilon(2)+\sigma_\epsilon(3)$, and the next one,
$\sigma_\epsilon(2+)$, is
$\sigma_\epsilon(2)+\sigma_\epsilon(3)+\sigma_\epsilon(4)$.)  Double
images dominate at low redshift, but triple and quadruple images are
increasingly important at higher redshift.  Due to the low
amplification bias, they never dominate, however.  Accounting for
BATSE's detection efficiency, we show the enhancement of the detection
rate, $N_\epsilon/N_0$, as the dashed line.  The elliptical potential
gives an increased detection rate of at most a few tens of per cent.
At redshifts $\zs\gsim 2.5$, fewer than 10/10000 sources in the Monte
Carlo simulation have observable multiple images, and numerical noise
dominates the results.

The caustic curves of an elliptical lens divide the source plane into
regions capable of producing one, three, and five images.  The five
image region is bounded by a diamond-shaped astroid, and the
associated cross section for producing five images is $\sigma_5$.  The
highest magnification events are associated with sources near folds or
cusps on the caustic.  The three image region is entirely outside this
astroid (as long as the core radius is small enough), and is bounded
on the outside by an ellipse with mean radius $r_E$.  The associated
cross section is $\sigma_3$, and bright images associated with appear
on opposite sides of the lens.  For lenses with small core radius, the
central image of a three or five image lens is highly demagnified, and
since we use a singular lens, the central image is always absent.
Thus, the $\sigma_3$ and $sigma_5$ cross sections produce two and four
images, respectively.

The elliptical lens produces fewer multiple imaging events with images
above the detector threshold than does the circular lens.  This is
essentially because, unlike a circular lens which produces high
magnification images in pairs, an elliptical lens can produce one (or
more) bright image of a source near a cusp, while leaving its
counter-image magnified much less (\eg Narayan \& Grossman 1989).
Highly magnified pairs or triplets of images are likely to be mergers
across folds and cusps.  Counter-images with lower magnification will
not be seen at redshifts far beyond $\zmax$, and indeed there are
essentially no quadruple images in Figure 4 beyond $\zs\approx 1.2$.

For a lens with $\epsilon=0.2$ and small core radius such that the
diamond caustic (defining the five-image region) lies entirely within
the the radial caustic (bounding the three-image region), the
three-image cross section $\sigma_3$ is greater than the five-image
cross section $\sigma_5$ at low and moderate magnifications
(Wallington \& Narayan 1993).  Images and counter-images can be seen
if a source is at low redshift, and, therefore, pairs of images
dominate the multiple image cross section $\sigma(2+)$.  The
three-image cross section varies as $\sigma_3\sim M^{-2.5}$, while the
five-image cross section varies as $\sigma_5\sim M^{-2}$, so that at
large magnifications, the five-image cross section dominates over the
three-image cross section.  The transition to $\sigma_5$ dominance is
at magnification $M\sim 10$ (Wallington \& Narayan 1993).  Thus,
three- and four-image bursts become increasingly more important at
larger redshifts, where bursts must be magnified more just to be seen.
(We associate both three- and four-image bursts with $\sigma_5$ since
they arise mainly from cusp imaging.  Two-image bursts are mainly
associated with $\sigma_3$ when they are at low redshift.  At high
redshift, however, the double images are mergers across folds, rather
than image/counter-image pairs, so that these pairs are also
associated with $\sigma_5$.)  However, due to the low magnification
bias, very high magnification events are not very common, and three-
and four-image bursts are never seen to dominate over two-image
bursts, unlike quasar imaging.

Although the absolute rate of observable multiple imaging by an
elliptical potential is reduced by factor
$\sigma_\epsilon(2+)/\sigma_0(2)$, it is possible that the rate at
which lensing events are detected is increased.  The enhancement of
the rate at which lensing events are detected is given by the multiple
image cross sections, weighted by the BATSE detection probability,
according to
$${N_\epsilon\over{N_0}}={0.12\sigma_\epsilon(2)+0.27\sigma_\epsilon(3)
+0.42\sigma_\epsilon(4)\over{0.12\sigma_0(2)}}.\putnum$$ This ratio is
drawn as the dashed line in Figure 4.  At the low and intermediate
redshifts that give quadruple bursts, the detection rate is enhanced
above the circular lens rate.  For the most probable source redshift
of $\zs\approx 0.8$ (see Fig. 2, $\Omega=1$, $\Lambda=0$ curve), the
detection rate is about 30\% larger than for a circular lens.  If
$\epsilon=0.1$, this enhancement is about 15\%, and if $\epsilon=0.3$,
it is about 70\%.  Averaged over all source redshifts, the enhancement
of the lensing rate clearly is less.  Thus, BATSE's efficiency to
detect lensing events is only slightly better than 12\%, being perhaps
around 15\%.

In Figure 5 we show the time delay distributions for the two Monte
Carlo simulations presented in Figure 4.  The $\epsilon=0$
distribution closely resembles the $\Omega=1$, $\Lambda=0$ model in
Figure 3, except for the absence of a divergence at small $\Delta t$,
resulting from the absence of lenses with $L<0.1L_*$ in the
simulation.  The $\epsilon=0.2$ distribution shows many more short
time delay bursts, resulting from images merging across folds.  (This
distribution includes all visible pairs in a multiple image burst.
For example, a quadruple burst has six time delays.)  The median time
delay in this distribution is about three times shorter than in the
$\epsilon=0$ distribution.  (Both medians here are systematically too
large due to the absence of low mass lenses, but if we calibrate the
$\epsilon=0$ median with the $0.02\,h_{100}^{-1}$ year interval
computed from the analytic model, we can get an idea of the true
median of the $\epsilon=0.2$ distribution.)  Thus, the time delays
associated with triple and quadruple imaging may often be less than a
day, and most often will be associated with the faintest bursts which
are seen only at high magnification.

\fig[PS/tdistm,\hsize,.6\hsize] 5.
This distribution of time delays between pairs of images derived from
the Monte Carlo simulation for $\epsilon=0$ (solid histogram) and
$\epsilon=0.2$ (dashed histogram).  The curves should be compared to
the $\Omega=1$, $\Lambda=0$ model in Fig. 3.  The curves do not
diverge at small $\Delta t$ because of the absence of lensing galaxies
fainter than $L=0.1L_*$ in the simulation.  Images merging across
folds produce many more short time delays in the $\epsilon=0.2$ model
than in the $\epsilon=0$ model, giving a median time delay about three
times shorter.

\section Discussion and Conclusion

We have presented three new cosmological models of the GRB population,
derived using a double power-law model of burst spectra (TM93), and
have computed the rate of double imaging of GRBs by the known
population of galaxies, using three different models of galaxy lenses.
The highest lensing rate is derived assuming a large cosmological
constant and a singular isothermal lens, while the lowest rate has
zero cosmological constant and a lens with finite core radius.  We
predict there should be one lensing event above the BATSE detection
threshold every 1.5--25 years.  The median time delay between the pair
of images is $\sim 0.02$ years.  Unfortunately, the instrumental
inefficiency associated with BATSE's duty cycle (12\% probability of
detecting both components of a double burst) is so severe that seeing
a lensing event from a galaxy lens is highly unlikely.

The instrumental inefficiency of observing a lensing event is improved
if four components of a burst, such as produced by an elliptical
potential, are above the BATSE threshold.  In this case there is a
42\% chance of detecting at least two of the four images.  Monte Carlo
simulations of lensing by elliptical potentials show, however, that
quadruple imaging is less common than for quasar lensing, essentially
because the amplification bias of lensed bursts is smaller.
Consequently, the rate at which lensed bursts will be detected is
increased by at most about 20\%.

Despite our pessimistic prediction for the detection rate of lensing
events, we believe it is well worth the effort to study the BATSE
light curves for possible lensing events.  A single lensing event,
albeit improbable, would verify the cosmological origin of GRBs.
Furthermore, even a single detection would suggest one of two exciting
possibilities.  It might indicate that the universe has a nonzero
cosmological constant, giving a larger lensing rate than the
``standard'' ($\Omega=1$, $\Lambda=0$) model.  An alternative
explanation could be the existence of a significant cosmological
density of compact massive lenses (\eg $10^6M_\odot$ black holes), in
which case the lensing rate could be much higher than for galaxy
lensing (Mao 1992b; Blaes \& Webster 1992).  The absence of lensed
bursts can be used to set constraints on the density of such massive
compact objects (Nemiroff \etal 1993).

Nowak and Grossman (1994) have proposed statistical tests for
comparing noisy light curves and find that many bursts of comparable
duration, but distinct origin can be distinguished by their methods.
Their methods are extremely reliable for the 41\% of bursts with
$\phi/\phi_{\rm min}\ge 3$, but for fainter bursts noise often will be
sufficient to obscure the differences in the light curves.  Thus,
consistent pairs of light curves among the brighter bursts would be
good lensing candidates.  Of course, one can only demonstrate that two
bursts are different, but not that they have the same source.  The
frequency of {\it false positives}, that is, error measures that are
consistent with the lensing hypothesis for two bursts that are
actually different, could be investigated by applying the statistical
tests to bursts which are known to be distinct (\eg because they
appeared in different parts of the sky).  If false positives are very
rare, then a pair of consistent light curves may constitute strong
evidence of lensing.  Calibration of the false positive rate has not
been undertaken, however, due to the enormous number of comparisons
(thousands) involved.  Nowak and Grossman have examined 260 bursts
preceding March 5, 1992 in the BATSE catalogue (Fishman \etal 1992),
and identified 27 pairs of bursts of comparable duration whose
positional error boxes overlap.  As expected, there are no candidate
lensing events among the bright ($\phi/\phi_{\rm min}$) bursts, and
only one fainter statistically consistent pair.  However, the duration
of this pair of bursts is so short that there are not enough time bins
in the light curves for high quality statistical comparison.  Thus, to
date there are no good lensing candidates in the data.

We emphasize that the improbability of detecting a lensing event is
not physical, but mainly instrumental.  An experiment designed with
lensing in mind could greatly increase the odds of a successful
lensing detection.  Most obviously, a satellite in a higher orbit
could have a duty cycle near 100\%, both because of smaller Earth
blockage and because the detectors could operate continuously.  Then,
instead of 12\% instrumental efficiency to detect both components of a
lensed burst, there would be nearly 100\% instrumental efficiency.  We
also note that three BATSE detectors in a comparable low orbit could
also give 100\% efficiency (almost a ten-fold increase in efficiency
for three times the cost!).

TM93 have proposed that a ``BATSE detector'' with 18 times greater
sensitivity could detect enough distant bursts in three years to
distinguish between $q_0=0.1$ and $q_0=0.5$ cosmologies.  In this case
the lensing rate would be about 4 times larger, giving one lensed
burst every 0.4--6 years, with 2.7 years being the mean for the
$\Omega=1$, $\Lambda=0$ cosmology.  If GRBs are cosmological in
origin, an appropriately designed experiment with high instrumental
efficiency would probably detect lensing events.  Failure to detect
any candidate lensing events within 4.6 times the mean burst interval
(say $4.6\times 2.7\approx 12$ years) would exclude the cosmological
scenario at 99\% confidence.

\bigskip

The authors are indebted to Man Hoi Lee and Chris Thompson for
discussions and for pointing out certain references during the course
of this project.
\frenchspacing

\maybebreak{.2} \bigskip\bigskip\bigskip
\centerline{\eightrm REFERENCES} \mark{References} \smallskip

\rj{Band, D. 1992}{\apj}{400}{L63}
\rj{Band, D., Matteson, J., Ford, L., Schaefer, B., Palmer, C.,
Teegarden, B., Cline, T., Briggs, M., Paciesas, W., Pendleton, G.,
Fishman, G., Kouveliotou, Meegan, C., Wilson, R., \& Lestrade, P.
1993}{\apj}{413}{281}
\rj{Blaes, O.M. 1994}{\apjs}{in press}{}
\rj{Blaes, O.M. \& Webster, R.L. 1992}{\apj}{391}{L63}
\rj{Blandford, R.D. \& Narayan, R. 1992}{\annrev}{30}{311}
\rj{Carrol, S.M., Press, W.H., \& Turner, E.L. 1992}{\annrev}{30}{499}
\rj{Clarke, T.E., Blaes, O., \& Gremaine, S. 1994}{\aj}{submitted}{}
\rj{Dermer, C.D. 1992}{Phys. Rev. Let.}{68}{1799}
\rj{Dyer, C.C. \& Roeder, R.C. 1973}{\apj}{180}{L31}
\rj{Fenimore, E.E., Epstein, R.I., Ho, C., Klebesadel, R. W., Lacey, C.
Laros, J.G. Meier, M., Strohmayer, T., Pendleton, G., Fishman, G.,
Kouvelioutou, C., \& Meegan, C. 1993}{\nat}{366}{40}
\rj{Fishman, G.J., \etal 1992}{BATSE Burst Catalog, GROSSC}{}{}
\rj{Fukugita, M., Futamase, T. \& Kasai, M. 1990}{\mnras}{246}{24p}
\rj{Fukugita, M., Futamase, T., Kasai, M., \& Turner, E.L. 1992}{\apj}{393}{3}
\rj{Fukugita, M. \& Turner, E.L. 1991}{\mnras}{253}{99 (FT91)}
\rconf{Gehrels, N., Chipman, E., \& Kniffen, D.A. 1993}{Compton Gamma-Ray
Observatory}{M. Friedlander, N. Gehrels, and D.J. Macomb}{American
Institute of Physics}{3}
\rj{Gould, A. 1992}{\apj}{386}{L5}
\rj{Grossman, S.A. \& Saha, P. 1994}{\apj}{submitted}{}
\rj{Hinshaw, G. \& Krauss, L.M. 1987}{\apj}{320}{468}
\rj{Huchra, J., Davis, M., Latham, D., Tonry, J. 1983}{\apjs}{52}{89}
\rj{Lauer, T.R. 1985}{\apj}{292}{104}
\rj{Lubin, L.M. \& Wijers, R.A.M.J. 1993}{\apj}{in press}{}
\rj{Mao, S. 1992a}{\apj}{380}{9}
\rj{Mao, S. 1992b}{\apj}{389}{L41}
\rj{Mao, S. 1993}{\apj}{402}{382}
\rj{Mao, S. \& Paczy\'nski, B. 1992a}{\apj}{388}{L45}
\rj{Mao, S. \& Paczy\'nski, B. 1992b}{\apj}{389}{L13}
\rj{Maoz, D. \& Rix, H.-W. 1993}{\apj}{416}{425}
\rj{Maoz, E. 1993}{\apj}{414}{877}
\rj{Meegan, C.A., Fishman, G.J., Wilson, R.B., Paciesas, W.S.,
Pendleton, G.N.  Horrack, J.M., Brock, M.N., \& Kouvelioutou, C.
1992}{\nat}{355}{143}
\rconf{Narayan, R. and Wallington, R. 1992}{Gravitational Lenses}{R. Kayser,
T. Schramm, and L. Nieser}{Springer-Verlag}{12}
\rconf{Narayan, R. and Grossman, S.A. 1989}{Gravitational Lenses}
{J.M. Moran, J.N. Hewitt, K.Y. Lo}{Springer-Verlag}{31}
\rj{Narayan, R. and Wallington, R. 1992a}{\apj}{399}{368}
\rconf{Narayan, R. and Wallington, R. 1992b}{Gravitational Lenses}
{R. Kayser, T. Schramm, L. Nieser}{Springer-Verlag}{12}
\rj{Nemiroff, R.J., Norris, J.P., Wickramasinghe, W.A.D.T., Horrack, J.M.,
Kouveliotou, C., Fishman, G.J., Meegan, C.A., Wilson, R.B., \&
Paciesas, W.S.  1993}{\apj}{414}{36}
\rj{Norris, J.P., Nemiroff, R.J., Scargle, J.D., Kouvelioutou, C.,
Fishman, G.J., Meegan, C.A., Paciesas, W.S., \& Bonnell, J.T. 1994}{\apj}
{in press}{}
\rj{Nowak, M.A. \& Grossman, S.A. 1994}{\apj}{submitted}{}
\rj{Ostriker, J.P. \& Vietri, M. 1986}{\apj}{300}{68}
\rj{Paczy\'nski, B. 1986}{\apj}{308}{L43}
\rj{Paczy\'nski, B. 1987}{\apj}{317}{L51}
\rj{Petrosian, V. 1993}{\apj}{402}{L33}
\rj{Piran, T. 1992}{\apj}{389}{L45}
\rj{Sanitt, N. 1976}{\mnras}{174}{91}
\rj{Sasaki, S. \& Takahara, F. 1993}{\mnras}{262}{681}
\rj{Schmidt, M. 1968}{\apj}{151}{393}
\rj{Tamblyn, P. \& Melia, F. 1993}{\apj}{417}{L21 (TM93)}
\rj{Turner, E.L. 1990}{\apj}{365}{L43}
\rj{Turner, E.L., Ostriker, J.P., \& Gott, J.R. 1984}{\apj}{284}{1}
\rj{Wallington, S. and Narayan, R. 1993}{\apj}{403}{517}
\rj{Wambsganss, J. 1993}{\apj}{406}{29}
\rj{Wickramasinghe, W.A.D.T., Nemiroff, R.J., Norris, J.P., Kouvelioutou, C.,
Fishman, G.J., Meegan, C.A., Wilson, R.B., \& Paciesas, W.S.
1993}{\apj} {411}{L55}

\nonfrenchspacing
\setbox\strutbox=\hbox{\vrule
        height9pt depth2pt width0pt}

\vfill\eject

\vbox{\centerline{TABLE 1}
\centerline{Cosmological GRB Models}
\smallskip
\centerline{
\vbox{\offinterlineskip\halign{
 \strut#\hfil\qquad&#\hfil\quad\quad&$#$\hfil\quad\quad&$#$\hfil\quad\quad
&$#$\hfil\cr
\noalign{\hrule\smallskip}\noalign{\hrule\smallskip}
$\Omega$&$\Lambda$&L_0h_{100}^2 \rm erg\,s^{-1}&n_0h_{100}^{-3}
\rm yr^{-1}\,Mpc^{-3}&z_{\rm max}\cr
\noalign{\smallskip\hrule\smallskip}
1.0 & 0.0 & 1.1\times 10^{50} & 7.8\times 10^{-8} & 0.83 \cr
0.1 & 0.0 & 2.0\times 10^{50} & 4.2\times 10^{-8} & 0.94 \cr
0.1 & 0.9 & 6.0\times 10^{50} & 1.3\times 10^{-8} & 1.13 \cr
}}}}
\setbox\strutbox=\hbox{\vrule
        height9pt depth2pt width0pt}
\catcode`\@=\active \def@{\hphantom{0}}

\vskip5.truecm

\vbox{\centerline{TABLE 2}
\centerline{Cumulative Lensing Rates and Amplification Biases}
\smallskip
\centerline{
\vbox{\offinterlineskip\halign{
 \strut#\hfil\qquad&#\hfil\quad
	&\hfil#\hfil\quad&\hfil#\hfil\quad\quad
	&\hfil#\hfil\quad&\hfil#\hfil\quad\quad
	&\hfil#\hfil\quad&\hfil#\hfil\cr
\noalign{\hrule\smallskip}\noalign{\hrule\smallskip}
\multispan2{\hfil}&\multispan2\hfil SIS\hfil\quad
	   &\multispan2\hfil NSIS\hfil\quad
	   &\multispan2\hfil QDEV\hfil\quad\cr
$\Omega$&$\Lambda$&$N_l\rm (yr^{-1})$&$B$&$N_l\rm (yr^{-1})$&
$B$&$N_l\rm (yr^{-1})$&$B$\cr
\noalign{\smallskip\hrule\smallskip}
1.0 & 0.0 & 0.09 &1.02& 0.04 & 1.91 & 0.04 & 0.01@@ \cr
0.1 & 0.0 & 0.17 &0.98& 0.08 & 0.48 & 0.07 & 0.0005 \cr
0.1 & 0.9 & 0.67 &9.94& 0.35 & 1.56 & 0.20 & 0.0004 \cr
}}}}

\closeout1 \vfill\eject
               \leftline{\bf\uppercase{figure captions}}
               \openout1=pscalls \input captions \closeout1
               \vfill\supereject \input pscalls \end